\documentclass{article}

\usepackage[numbers]{natbib}         
\usepackage[colorlinks]{hyperref}    
\usepackage[english]{babel} 
\usepackage{amssymb}
\usepackage{amsmath}
\usepackage{txfonts}
\usepackage{mathdots}
\usepackage[classicReIm]{kpfonts}
\usepackage[dvips]{graphicx} 
\usepackage[a4paper, portrait, margin=1in]{geometry}
\usepackage{tabularx}
\usepackage{longtable}
\usepackage{multirow}
\usepackage{booktabs}
\usepackage[labelsep=period]{caption}
\usepackage{makecell}
\begin{document}
	\setlength{\parindent}{0pt}
	\setlength{\parskip}{1ex}
	
	\textbf{\Large Data-Driven Volumetric Image Generation from Surface Structures using a Patient-Specific Deep Leaning Model}
	
	\bigbreak

	Shaoyan Pan$^{1,2,5}$, Chih-Wei Chang$^{1,5}$, Marian Axente$^{1}$, Tonghe Wang$^{3}$, Joseph Shelton$^{1}$, 
	Tian Liu$^{4}$, Justin Roper$^{1}$ and Xiaofeng Yang$^{1,2*}$

	1Department of Radiation Oncology and Winship Cancer Institute, Emory University, Atlanta, GA 30308
	
	2Department of Biomedical Informatics, Emory University, Atlanta, GA 30308
	
	3Department of Medical Physics, Memorial Sloan Kettering Cancer Center, New York, NY, 10065
	
	4Department of Radiation Oncology, Mount Sinai Medical Center, New York, NY, 10029
	
	5These authors contributed equally: Shaoyan Pan, Chih-Wei Chang

	\bigbreak
	\bigbreak
	\bigbreak

	\textbf{*Corresponding author: }
	
	Xiaofeng Yang, PhD
	
	Department of Radiation Oncology
	
	Emory University School of Medicine
	
	1365 Clifton Road NE
	
	Atlanta, GA 30322
	
	E-mail: xiaofeng.yang@emory.edu

	\bigbreak
	\bigbreak
	\bigbreak
	\bigbreak
	\bigbreak
	\bigbreak

	\textbf{Abstract}

	The advent of computed tomography significantly improves patients’ health regarding diagnosis, prognosis, and treatment planning and verification. However, tomographic imaging escalates concomitant radiation doses to patients, inducing potential secondary cancer by 4\%. We demonstrate the feasibility of a data-driven approach to synthesize volumetric images using patients’ surface images, which can be obtained from a zero-dose surface imaging system. This study includes 500 computed tomography (CT) image sets from 50 patients. Compared to the ground truth CT, the synthetic images result in the evaluation metric values of 26.9 ± 4.1 Hounsfield units, 39.1 ± 1.0 dB, and 0.965 ± 0.011 regarding the mean absolute error, peak signal-to-noise ratio, and structural similarity index measure. This approach provides a data integration solution that can potentially enable real-time imaging, which is free of radiation-induced risk and could be applied to image-guided medical procedures.
	
	\bigbreak
	\bigbreak

	\noindent 
	\section{ INTRODUCTION}
	
	The advent of computed tomography (CT) has caused a paradigm shift in the field of radiation oncology regarding qualitative lesion diagnosis and prognosis, quantitative treatment planning, and precise radiotherapy delivery. Tomography leverages enormous x-ray projections with superimposed anatomical details collected from various detector angles to reconstruct three-dimensional (3D) images for a patient. Analytical and iterative image reconstruction methods have been developed along with compressed sensing techniques\cite{RN733, RN734, RN735} to reconstruct CT images efficiently and effectively. Recently, deep learning (DL)\cite{RN533} methods have been proposed to increase the quality of images while accepting less x-ray projections. However, each x-ray projection imparts ionizing radiation to the patient, which raises concern for patient safety due to increased radiation dose by acquiring sufficient x-ray projections (Fig. 1a). Due to the need of radiotherapy, conventional clinical workflows can potentially increase the total patient dose by 2\%\cite{RN702, RN703}, increasing the probability for secondary cancer by 4\%\cite{RN700}. In contrast, optical surface imaging guidance systems\cite{RN741, RN740} has been recently deployed in the clinic as a zero-dose alternative to transmission x-ray imaging in radiotherapy. Unfortunately, these systems can merely image the underlying volumetric patient anatomy. Unlike x-ray projections, the surface image contains zero-prior patient anatomical information. How to image patient with zero radiation dose remains an open question for advancing the tomographic paradigm.
	
	This work aims to investigate the feasibility of generating 3D anatomical images from a 3D surface image of a patient without using any patient anatomical prior. More specifically, a DL framework is designed to infer the hidden spatial anatomical details using the ultra-sparce body surface. The proposed framework can potentially provide real-time images to guide the delivery of photon and proton therapy and especially for FLASH radiotherapy\cite{RN554}, which requires treatment delivery with high accuracy. Such a data-driven approach maximally leverages prior knowledge to enhance image prediction by integrating data from different modalities, which is an emerging topic in multiple disciplines\cite{RN244, RN174}. Recent studies\cite{RN178, RN713, RN714, RN715} have shown that machine learning (ML) can achieve data-driven image reconstruction using sparse projections. While the projection still contains 2D line integrals of patient anatomy superimposed in a 2D domain, in this work, we demonstrate 3D CT synthesis via prior knowledge from patient-specific surface images, which lack the line integral anatomical information. Importantly, this new approach can increase patient safety by replacing ionizing x-ray imaging with non-ionization imaging and would enable continuous, real-time image guidance during radiation delivery, which is not currently possible using the available x-ray imaging systems.
	
	Traditional ML methods have succeeded in data-driven modeling due to their model simplicity\cite{RN179, RN180} regarding generalizability and interpretability, but the performance of these models can be compromised when using a substantial amount of data without sufficient knowledge\cite{RN256, RN244, RN279}. In contrast, deep neural networks with hierarchical model layers have been demonstrated as universal approximators\cite{RN183} capable of recognizing patterns and synthesizing images from complex data structures\cite{RN183}. Herein, we propose a generative adversarial, surface-to-volume network with hierarchical architecture to transform surface images from the patient-specific body structure to 3D synthetic CT. The model architecture includes a reconstruction network to extract the feature maps from surface images and transform those maps into the essential tensor feature maps for volumetric image reconstruction (Fig. 1b). Although the surface image does not include spatial details of patient anatomy, the variable features of the patient body surface can embed hidden functional relationships to patient CT images. During training phase, the network learns the underlying correlations between low-dimensional (surface image) and high-dimensional (volumetric image) feature maps to be able to later generate synthetic CT datasets based on patient-specific body surface structure. A proposed refinement network is designed to ensure the consistency between the synthetic and actual images acquired from the clinically commissioned CT scanner, calibrated for radiotherapy simulations as part of each patient treatment planning. The refinement network is trained to learn the intensity distribution from actual CT images to conserve the CT noise level and contrast resolution. Since image synthesis from surface images to volumetric images is inherently ill-posed, prior knowledge from the specific patient can therefore effectively inform DL networks to generate 3D synthetic CT using merely body structure from patients. 
	
	\begin{figure}
		\centering
		\noindent \includegraphics*[width=6.50in, height=4.20in, keepaspectratio=true]{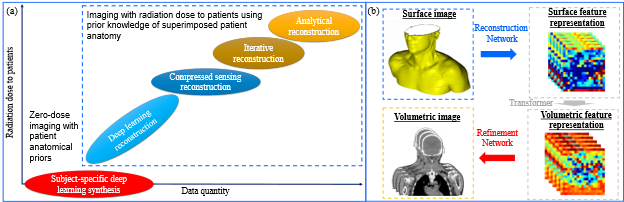}
		
		\noindent Fig. 1. Volumetric image reconstruction using x-ray projections and non-ionizing radiation surface images. a, Mechanical- and deep learning-based image reconstruction schemes in the context of radiation dose and data quantity. b, Volumetric image reconstruction using a surface image.
	\end{figure}
	
	\noindent 
	\section{Methods and Materials}
	\noindent 
	\subsection{Problem formulation}
	
	We formulate the proposed surface-to-volume CT reconstruction as a 3D volume-to-volume translation problem. We first represent a human body surface in a 3D matrix $X \in R^{(H\times W\times L)}$ where H, W, and L are lengths of the matrix's axial, coronal, and sagittal axis. In this surface matrix, only the voxels along the surface preserve the original Hounsfield Unit (HU) intensities. The voxel intensities of all the inner structures and organs are set to the background's intensities. Then we aim to translate the inner intensities to actual intensities describing the corresponding 3D anatomy scene, in a voxel-wise manner, by relying only on the surface's intensities. Formally, with the surface matrix X as input, the surface-to-volume framework $f_\theta$ with parameter $\theta$ outputs a complete volume matrix $Y_{pre} \in R^{(H\times W\times L)}$:
	
	\begin{equation} 
		Y_{pre}=f_\theta{(X)}
	\end{equation} 
	
	The deep learning framework $f_\theta$ should be able to infer the volume matrix $Y_{pre}$ as close as possible to the ground truth volume matrix $Y_{true}$. The framework $f_\theta$ is therefore formulated as a minimizer of the objective function:
	\begin{equation} 
		\arg \min_\theta d(Y_{pre},Y_{true})
	\end{equation} 
	where $d$ is a measure of the similarity between the output $Y_{pre}$ and the target volume $Y_{true}$. 
	
	\noindent 
	\subsection{Volumetric image generation framework}
	
	As illustrated in Fig. 2, the proposed deep learning framework $f_\theta$ consists of a reconstruction, verification, and refinement network. We first create down-sampled versions for both the boundary input X and the target volume $Y_{true}$ as $X^{down}$ and $Y^{down}$, respectively. In the coarse reconstruction stage, a Cycle-consistent reconstruction-verification system is deployed to translate $X^{down}$ into a low-resolution body volume $Y_{pre}^{down}$. The reconstruction-verification system is motivated by cycle-consistent generative adversarial systems\cite{RN202}. Intuitively, without loss of information, an image translated from a source domain to a target should be able to translate from the target domain back to the source domain and be identical to the original image. Based on this idea, cycle-consistent architectures enhance the image translation ability by minimizing information loss and demonstrating state-of-the-art performance in many image-to-image translation tasks. Motivated by this idea, we design a cycle-consistent system consisting of reconstruction and verification networks, and reconstruction and verifications discriminators. The reconstruction network is used to translate the$X^{down}$ to$Y_{pre}^{down}$. The verification network is utilized to translated $Y_{pre}^{down}$ back to $X_{pre}^{down}$, which serves as a regularization to the reconstruction network, to minimize the information loss between the translations. In addition,  In the meantime, following the idea of generative adversarial networks\cite{RN90}, a reconstruction and verification discriminator are deployed to measure a Wasserstein distance\cite{RN91} between the $X_{pre}^{down}$ and $X^{down}$, and the $Y_{pre}^{down}$ and $Y^{down}$, respectively. The Wasserstein distances are an additional optimization target for the reconstruction and verification network so they can output images on their corresponding domains. 
	Then the $Y_{pre}^{down}$ is up-sampled to the size of the original target $Y_{true}$ by trilinear interpolation. Finally, in the refine reconstruction stage, the up-sampled $Y_{pre}^{down}$, which is denoted as $Y_{pre}^{coarse}$, is forwarded into an encoder-decoder network, to generate a fined volume reconstruction $Y_{pre}$. Such a multi-stage design utilizing the down-sampled data is necessary to output a high-resolution, detailed body volume under a memory constraint. In summary, following the proposed pipeline, we first generate a coarse low-resolution volume by the reconstruction-verification network to avoid overloading memory. Then it is simple but adequate for the refinement network to refine the high-resolution details on the up-sampled coarse volume.
	
	\noindent 
	\subsection{Reconstruction/verification network}
	
	The proposed reconstruction/verification networks are designed as an encoder-transformer-decoder network. The encoder down-samples the input sequentially to obtain multi-scale semantic features. The transformer will refine the semantic features further and forward them to the decoder. The decoder is a symmetric expanding path that decompresses the encoder’s features into the reconstructed volume. The proposed convolutional layers of the generators are designed based on the residual networks\cite{RN203} to help the network figure out hidden correlations from data via hierarchical structures. In our design, the reconstruction and verification networks have identical architecture. Here we describe the details of the encoder and the decoder.
	The encoder consists of an early convolutional layer and two down-sampling convolutional layers. To be concise, we present the kernel, stride, and padding sizes, which are isotropic in three dimensions by a single scalar. The early convolutional layers consist of a pattern of “3D convolutional layer (kernel size 3, stride size 1, and padding size 1) → 3D instance normalization (IN) layer\cite{RN81} → Rectified linear unit (RELU) activation\cite{RN290}”. The down-sampling layers have an almost identical path to the early convolutional layer, while the down-sampling layers deploy a convolutional layer with a stride size of 2 instead of 1. In practice, we present the size of the boundary matrix (1×80×128×80) and the three following convolutional blocks as (80×128×80→ 40×64×40→ 20×32×20) with 32, 64, and 128 filters at each block.
	The transformer consists of three residual convolutional blocks (RCBs). Each residual convolution block follows a "3D convolutional layer (kernel size 3, stride size 1, padding size 1 and with bias) → 3D IN layer → RELU activation → 3D convolutional layer (kernel size 3, stride size 1, padding size 1 and with bias) → 3D IN layer". Following the residual connection design, the convolutional path's output is added to the input to obtain the final output. Accordingly, the convolutional blocks' outputs are 20×32×20 with 128 filters at each block.
	The architecture of the proposed decoder is symmetric with the encoder in a reverse direction: two up-sampling and a sequential convolutional layer. An additional final convolutional layer is deployed at the end to output a reconstructed volume/surface. Each up-sampled convolutional block is formed by a “3D deconvolutional layer (kernel size 3, stride size 2, padding size 1, output padding size 1 and with bias) 3D (IN) layer Rectified linear unit (RELU) activation”. The sequential convolutional block follows the structure of the early convolutional block in the encoder. We present the changes in the feature maps' size and the final output of the decoder as 20×32×20→40×64×40→80×128×80→80×128×80→80×128×80 with 64, 32, 16, and 1 filters in the three convolutional blocks and the final convolutional block, respectively.
	
	\noindent 
	\subsection{Reconstruction/verification discriminators}
	
	The discriminators are built as four sequential down-sampling convolutional layers and one final convolutional layer. The down-sampling layers follows a pattern of “3D convolutional layer (kernel size 4, stride size 2, and padding size 1) → 3D IN layer → RELU activation”. The final convolutional layer has the same pattern, expect for the stride size of the convolutional layer is 1. To train the discriminators to measure a Wasserstein distance between the predicted volumes and the ground truth volumes, we connect linear regression module consisting of “global pooling layer → flatten → linear layer (output dimension is 1)” to the end of the last down-sampling block. The number of the filters of the four down-sampling convolution and final convolution layer are 64,128,256,512, and 512, respectively. Here we present the size of the features from the input volume to the final output: 80×128×80→40×64×40→ 20×32×20→10×16×10→5×8×5→5×8×5→1×1×1→1. 
	Accordingly, the reconstruction discriminator can output a scalar to represent the Wasserstein distance between the $Y_{pre}^{down}$and $Y_{true}^{down}$, and the verification network can output the distance between the $X_{pre}^{down}$ and $X^{down}$.
	
	\noindent 
	\subsection{Refinement network}
	
	By the reconstruction network, we can obtain the coarse reconstructed volume$Y_{pre}^{down}$. Then we obtain the corresponding blurry up-sampled $Y_{pre}^{coarse}\in R^{(160\times 256\times 160)}$. We propose a subsequential refinement network, which is another encoder-decoder network, to refine the high-resolution details on the blurry $Y_{pre}^{coarse}$. The architecture and the hyper-parameters of the refinement network are determined by the size and voxel spacing of the input scans, following the nnUnet’s principle. With the size of 160×256×160 and an isotropic spacing of 1 millimeter (mm) of the input, the refinement network’s encoder consists of one residual convolutional block (RCB), five down-sampling RCBs (D-RCBs), and a bottleneck residual convolutional block (B-RCB). The D-RCB and B-RCB have identical structure with the RCB, while the former replaces the stride size of the first 3D convolutional layer from 1 to isotropic size 2, and the latter revise the stride of the first 3D convolutional layer from isotropic size 1 to the anisotropic size 1×2×1. The numbers of filters are 32, 64, 128, 256, 320, 320 for the first RCB to the last B-RCB.  In addition, the sizes of the input and down-sampled features through the encoder are: 160×256×160→160×256×160→80×128×80→ ×40×64×40→ 20×32×20→10×16×10→ 5×8×5→5×4×5.  
	The refinement network decoder, which is symmetrical to its encoder, consists of one up-sampling bottleneck RCB (UB-RCB) and five up-sampling RCBs (U-RCB) and a final convolutional block. The UB-RCB and U-RCB deploy a “3D deconvolutional layer (kernel size 3×2×3, stride size 1×2×1 and padding size 1×0×1)” and a “3D deconvolutional layer (kernel size isotropic 2, stride size isotropic 2 and padding size isotropic 0)” to replace the first convolutional layer of the B-RCB and the D-RCB, respectively. The numbers of the filters of the deconvolutional layer equals to the numbers of the channels of the input features. The final convolutional block is formed by a 3D convolutional layer (kernel size 1, stride size 1, padding size 0 and without bias) with a Tanh activation. For the first to the sixth D-RCB and the final convolutional layer, the numbers of the filters are 320, 320, 256, 128, 64, 32, and 1, respectively. Additionally, the size of the input features and the up-sampled features through the decoder are: 5×4×5→ 5×8×5→ 10×16×10→ 20×32×20→40×64×40→ 80×128×80→160×256×160→160×256×160.
	
	\noindent 
	\subsection{Training procedure and objective function}
	
	The proposed surface-volume reconstruction pipeline is trained to translate the X to $Y_{true}$ in 2000 epochs. In each epoch, we firstly trained the coarse reconstruction stage throughout the whole training dataset with a mini-batch-size of 2, then trained the refined reconstruction stage throughout the dataset which is shuffled with a mini-batch-size of 1. To be concise, we present the training procedure in each epoch by three steps.
	Step 1: Training the reconstruction/verification networks. In the coarse reconstruction stage, we firstly optimize the reconstruction/verification networks in a cycle-consistent way: an input $X^{down}$ can be translated to $Y^{down}$ by the reconstruction network (denoted as $G_R$) then be brought back to $X^{down}$ by the verification network (denoted as $G_V$). And an input $Y^{down}$ can be translated to $X^{down}$ then be brought back to $Y^{down}$ if bot networks are well-functioning in their translation domains. To be more detailed, we express the $X_{pre}^{down}= G_V (Y^{down})$ and  $Y_{pre}^{down}= G_R (X^{down})$, and:
	
	\begin{equation} 
		L_{surf}= 0.5*MAE(G_V (Y^{down} ),X^{down} )+0.5*MAE(G_V (G_R (X^{down} )),X^{down} )
	\end{equation}

	\begin{equation} 
		L_{vol}= 0.5*MAE(G_R (X^{down} ),Y^{down} )+0.5*MAE(G_R (G_V (Y^{down} )),Y^{down} )
	\end{equation}

	where MAE is the mean absolute error. Then, the network should be jointly optimized to reduce the Wasserstein distance of the reconstruction/verification discriminators (denoted as $D_R$ and $D_V$), which can be simplified as: 
	
	\begin{equation} 
		L_{surf}= 0.5*MAE(G_V (Y^{down} ),X^{down} )+0.5*MAE(G_V (G_R (X^{down} )),X^{down} )
	\end{equation}

	Furthermore, it is helpful to encourage the reconstruction/verification networks to preserve high-level volume characteristics between the input and output by the identity loss\cite{RN202}. We deployed the loss to enforce the reconstruction and verification network to output identical matrix to the $X^{down}$ and the $Y_{true}^{down}$, respectively:
	
	\begin{equation}
		L_{identity}= 0.5*MAE(G_R (Y_{pre}^{down} ),Y^{down} )+ 0.5*MAE(G_V (X^{down} ),X^{down} )
	\end{equation}

	The final objective function for the reconstruction/verification networks is:
	
	\begin{equation}
		L= L_{surf}+L_{vol}+ L_{asserstein}+L_{identity}
	\end{equation}

	Step 2: Training the reconstruction/verification discriminators. After the optimization of the reconstruction/verification networks, we optimize the discriminators $D_R$ and $D_v$ to output Wasserstein distance between the $Y_{pre}^{down}$ (which is $G_V (Y^{down} )$)  and $Y^{down}$, the $X_{pre}^{down}$  (which is $G_V (Y^{down} )$) and $X^{down}$, respectively. To achieve that, we train them as by Wasserstein loss with gradient penalty:
	\begin{equation}
		L_{D-vol}= D_R (G_R (X^{down} ))- D_R (Y^{down} )+\lambda*(|(|\nabla_( Y_{mix} ) (D_R (Y_{mix} ))|)|_2-1)^2 
	\end{equation}

	\begin{equation}
		L_{D-suf}= D_V (G_V (Y^{down} ))- D_V (X^{down} )+\lambda*(|(|\nabla_( X_{mix} ) (D_V (X_{mix} ))|)|_2-1)^2 
	\end{equation}
	Where$||\cdot||_2$  indicates L2-norm, $\lambda$ is a hyperparameter empirically set to 10, $Y_{mix}$ is a random mixture uniformly from the pairs of $Y^{down}$ and $Y_{pre}^{down}$, $X_{mix}$ is a random mixture uniformly from the pairs of $X^{down}$ and $X_{pre}^{down}$, $\nabla_( Y_{mix} )$ is the $D_R$ and $D_v$’s gradients with respect to $Y_{mix}$ and $X_{mix}$, respectively Specifically, $Y_{mix}=\epsilon*Y_{pre}^{down}+(1-\epsilon*Y^{down} )$, where $\epsilon$ is random number sampled from a Beta distribution with two shape parameters equal to 0.2. 
	
	Step 3: Training the refinement network. After we train the reconstruction/verification system through the whole dataset, we can obtain the corresponding up-sampled reconstructed volumes. We start to optimize the refinement network to refine the sub-quality volumes $Y_{pre}$ by:
	
	\begin{equation}
		L_{refine}=MAE(Y_{pre},Y_{true})
	\end{equation}
	So far, we present all the training steps for each epoch. 
	
	\noindent 
	\subsection{Implementation details}
	The surface-to-volume pipeline was implemented using the PyTorch framework\cite{RN193} in Python 3.8.11 on a workstation running Windows 11 with a single Nvidia RTX 6000 GPU with 48GB memory. The Adam optimizer optimizes the networks with an initial learning rate of 10-4; the learning rate will decay to 0.93 of its current value for every 50 epochs. After every ten training epochs, the model was evaluated on the validation dataset to monitor the model performance on the “unseen” data. After the training of 2000 epochs, the models with the smallest refinement loss were saved as the final models. The training time of the network is around 19 hours; the inference time is around 13 seconds per volume.

	\noindent 
	\subsection{Material}
	The proposed network is evaluated by dataset of 50 patients, and each patient contains 4D CT collected from 10 respiratory phases. Independent experiments were conducted for each patient. All image data are acquired from Siemens SOMATOM Definition AS with a 120 kVp energy spectrum. The image resolution and voxel size are 512 x 512 x (133-168) and voxel spacing of 1.56 x 1.56 x 1.56 mm3 along the axial, coronal, and sagittal axis. In each experiment, the patient volume collected from the 70\% respiratory phase is used for inference; a volume from another random phase will be used for validation, and the remaining phases are used for training.
	
	\noindent 
	\subsection{Data preprocessing}
	The experimented volumes were firstly resampled to 3 x 3 x 1.56 mm3. All the volumes were then centered according to their center of mass. Then we cropped or zero-padded the boundary regions so that all volumes have a consistent resolution of 160 x 256 x 160.  For both the training and inference, the voxel intensities of volume and surface scans are normalized across entire dataset to the interval [0,1]. Data augmentations are conducted in the training stage to increase the diversity of the existing dataset to improve the network’s performance. In each epoch, each volume has a probability of 0.5 to be augmented by random volume shearing (with a random ratio of 0 to 0.1 in each axis), rotation (with a random angle range from 0° to 10° in each axis), and scaling (with random a ratio of 0 to 0.1 in each axis).
	
	\noindent 
	\subsection{Evaluation}
	We use three evaluation metrics to quantify the synthetic volumetric images, generated by the trained surface-to-volume network using testing datasets. The evaluation metrics are computed for all 50 patients (Fig. 3) and the mean values are summarized in Table 1. MAE is calculated within patients’ surface structures, and this metric quantifies absolute errors of CT numbers between the generated volumetric images and ground truth. SSIM\cite{RN522} and PSNR are used to measure the similarity and quality between the generated and ground-truth CT images.

	\noindent 
	\section{Results}
	
	Fig. 2 depicts the architecture of the surface-to-volume network. The patients’ body surface is the input, and the output is a 3D generated volumetric image (i.e., synthetic CT) corresponding to the input surface image. The surface structure is created using triangular meshes, while CT images are formatted in voxels. To construct the volumetric images, the surface structure is converted from meshes into voxels so that the directionalities are conserved between the surface and volumetric images\cite{RN742, RN743}. The network architecture is designed to include three modularized sub-networks: the reconstruction network, verification network, and refinement network. The reconstruction network first encodes the surface images into various feature maps and transforms the surface feature maps into volumetric feature maps, which allows the decoder to reconstruct the volumetric images. During the training phase, the network learns how to correlate the feature distributions between the surface and volumetric images. The verification network transforms the generated volumetric images back to generated surface images to ensure the invariance of DL-based image synthesis to the input surface images. Ultimately, the refinement network ensures the scale invariance of CT numbers between the output volumetric images and ground truth CT images. This network learns the high-density tissue intensity and image noise level and contrast from ground truth CT images, which is patient-specific and machine-specific prior knowledge. The implementation details of the surface-to-volume network are provided in Methods.
	
	\begin{figure}
		\centering
		\noindent \includegraphics*[width=6.50in, height=4.20in, keepaspectratio=true]{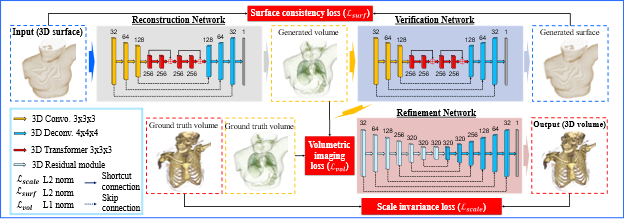}
		
		\noindent Fig. 2. Model hierarchy for the surface-to-volume network architecture. Model hierarchy of the deep learning networks including three primary components. The reconstruction network generates volumetric images from the surface image. The verification network manufactures a surface from the 3D synthetic CT to compare with the measured surface. The refinement network conserves the material attenuation characteristics between the ground truth CT images and synthetic CT images. The numbers denote the kernel sizes used in each layer within each network
	\end{figure}

	End-to-end DL training requires a substantial amount of data to achieve model robustness and predictive capability. In this work, the patient-specific prior knowledge consists of the four-dimensional (4D) CT simulation scan, which is commonly used when simulating lung patients to estimate and manage respiratory motion. 4DCT is typically formed from datasets binned based on various external signals into 3D CT datasets corresponding to ten respiratory phases. Five-hundred 3D CT image sets are used to generate surface images from fifty patients in the institutional database. Indeed, additional patient-specific CT images, such as quality assurance CT and daily/weekly cone-beam CT, can be available as requested by radiation oncologists. These images may be used to adapt the model and account for any changes in anatomy or respiration that can occur over the course of treatment.
	The surface-to-volume network can assimilate all the available, relevant, and adequately evaluated image data to improve the model performance. It should be mentioned that the surface-to-volume network is designed for patient-specific learning, and we train fifty network models (Fig. 2) for each patient. Data augmentation techniques, including translational and rotational sampling, are applied to CT image sets to forecast potential clinical scenarios. Details of model training and testing are described in Method.
	
	The model can be used for real-time, surface-to-volume image generation to visualize patient anatomy including tumor locations, allowing radiation oncologists to verify the intra-fractional motion. We explore the feasibility of reconstructing volumetric images from body surface structure using 4DCT from patients with lung cancer, as motion management is crucial to guide radiation to the moving target while sparing the nearby tissues to the extent possible – thereby increasing the therapeutic ratio. To evaluate all fifty surface-to-volume networks, we use three metrics to investigate under which conditions the volumetric images can be generated with minimum uncertainty. The mean absolute error (MAE) and peak signal-to-noise ratio (PSNR) are used to measure the quality of generated volumetric images regarding noise level, contrast, and CT numbers. The structural similarity index measure (SSIM)\cite{RN522} examines the similarity between two images, thus comparing the synthetic CT anatomy to the original simulation CT. Based on the last three-evaluation metrics, the k-mean clustering is used to classify the results of generated volumetric images into different groups to investigate under which conditions the model can predict ground truth CT with minimal uncertainty. It should be emphasized that the goal is to generate CT-like images, which can maximally reconstruct actual CT characteristics\cite{RN367, RN552} for radiation oncology treatment planning.
	
	Fig. 3 depicts the violin plots to classify the evaluation results of patient images in three groups using k-means clustering\cite{RN745, RN746} based on MAE, PSNR, and SSIM. The figure includes probability distributions and interquartile ranges to measure the performance of the proposed method. The mean values of each metric show that the surface-volume network can generate 3D CT images with minimum uncertainty for patients in group 1 (Table 1). Fig. 4 shows the reconstructed volumetric images and ground-truth CT in transversal views for different cases from different groups. Fig. 4 also includes comparisons of SSIM, difference maps, and CT-number line profiles between synthetic images and ground truth. Fig. 5 illustrates the sagittal views of the reconstructed volumetric images, together with the ground-truth CT and histogram comparisons of CT numbers between generated images and ground truth. All evaluation metrics indicate the model potential regarding generating comparative volumetric images to 3D CT images acquired from treatment planning CT scanners. Indeed, the results indicates that the performance of deep hierarchical networks differs for patients in different groups (Fig. 3-5). The surface image datasets with high quality should increase the predictive capability as well as the learning efficacy, which potentially makes the surface-to-volume model deployable in the clinic.
	
	\begin{figure}
		\centering
		\noindent \includegraphics*[width=6.50in, height=4.20in, keepaspectratio=true]{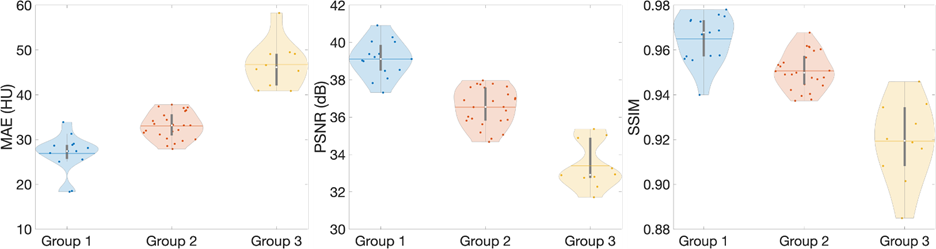}
		
		\noindent Fig. 3. Evaluate the performance of the surface-to-volume image reconstruction network. Violin plot to show the distributions of each evaluation metrics including mean absolute error (MAE), peak signal-to-noise ratio (PSNR), and structural similarity index measure (SSIM). Groups are classified using k-mean clustering with inputs from the three-evaluation metric.
	\end{figure}

	\begin{table}[]
		\centering
		\caption{Reconstruction results for the generated volumetric images from surface images. MAE, mean absolute error; PSNR, peak signal-to-noise ratio; SSIM, structural similarity index measure.}
		\label{tab:my-table}
		\resizebox{\textwidth}{!}{%
			\begin{tabular}{llll}
				\hline
				Group & MAE (HU) & PSNR (dB) & SSIM \\ \hline
				1 & 26.9 ± 4.1 & 39.1   ± 1.0 & 0.965   ± 0.011 \\ \hline
				2 & 33.1 ± 2.9 & 36.5   ± 1.0 & 0.951   ± 0.008 \\ \hline
				3 & 46.7 ± 5.2 & 33.4   ± 1.3 & 0.919   ± 0.018 \\ \hline
				Average & 34.0 ± 7.9 & 36.7   ± 2.3 & 0.949   ± 0.020 \\ \hline
			\end{tabular}%
		}
	\end{table}

	\begin{figure}

		\noindent \includegraphics*[width=6.50in, height=4.20in, keepaspectratio=true]{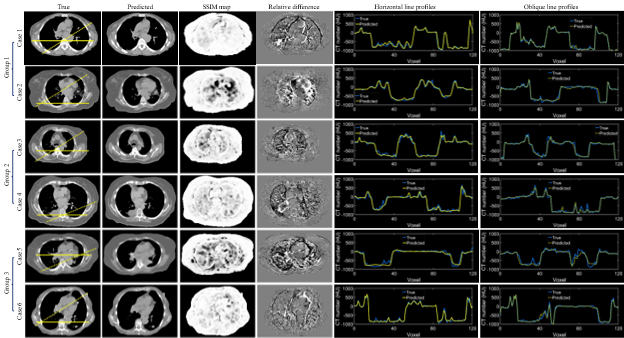}
		
		\noindent Fig. 4. Examples of predicted volumetric images from each group. The transversal views of ground truth and predicted CT are displayed. The evaluation metrics include structural similarity index measure (SSIM), relative difference maps, and line profiles. The horizontal solid and oblique dot lines on the ground truth images indicate the location of profile comparisons. The training, validation, and testing datasets for each case include 1280, 160, and 160 CT images.
	\end{figure}

	\begin{figure}

		\noindent \includegraphics*[width=6.50in, height=4.20in, keepaspectratio=true]{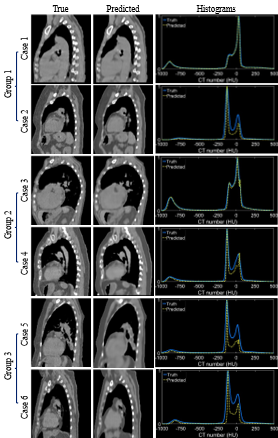}
		
		\noindent Fig. 5. | Examples of predicted volumetric images from each group with histogram distributions of CT numbers. The sagittal views of ground truth and predicted CT are displayed. The training, validation, and testing datasets for each case include 1280, 160, and 160 CT images.
	\end{figure}

	\bigbreak
	
	\noindent 
	\section{Discussion}
	
	To further explore the deep surface-to-volume network, we analyze the data quality of surface images from each classified group to determine the potential of the generated DL model regarding semantic cognitive reasoning. Fig. 6a depicts the t-distributed stochastic neighbor embedding (t-SNE)\cite{RN744} visualization of k-mean clustering\cite{RN745, RN746} results of the generated volumetric images for all fifty patients. The t-SNE is widely used to embed high-dimensional data in a 2D or 3D domain while conserving the object similarities based on probability. Fig. 6a shows that three cluster groups are well-differentiated, and generated image sets within the same group have similar metric values. Based on the clustering results, we compute the surface curvature\cite{RN742, RN743} for patients in each corresponding group. The coefficient of variance (COV) is used to measure the spread of the curvature distribution. The figure shows that mean COV increase from group 3 to group 1, which is consistent with the trend of evaluation metric change between groups. The result suggests a positive correlation between surface curvature and model performance. 
	Fig. 6b illustrates the coefficient of variance distributions of surface curvature\cite{RN742, RN743} in each group where the mean values are 0.932, 0.845, and 0.809. Fig. 6c shows the representative surface curvature map using the data randomly sampled from each group. The curvature includes different sign conventions due to the concave or convex nature of the surface property. However, we are only interested in the magnitude of the curvature since we hypothesize that complex surfaces should provide rich information, which can potentially embed more features for the proposed volumetric generation model to discover. The results indicate that the surface images with more curvature variation tend to reduce the learning barrier for the model such that the surface-to-volume network can generate volumetric images with less uncertainty. This finding suggests that the surface-to-volume method could benefit patients who received radiotherapy without clothing or masking since those artificial coverings cause biases for surface detection. 
	
	\begin{figure}
		
		\noindent \includegraphics*[width=6.50in, height=4.20in, keepaspectratio=true]{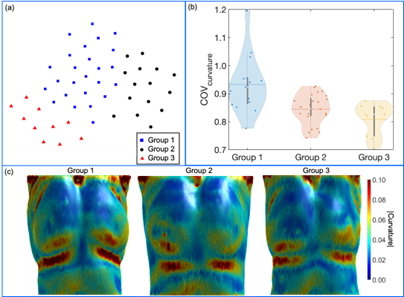}
		
		\noindent Fig. 6. | Analysis of generated volumetric images. a, The t-distributed stochastic neighbor embedding (t-SNE) visualization of the k-mean clustering results of the total generated volumetric images based on the evaluation metrics of MAE, PSNR, and SSIM. b, Violin plot to show the coefficient of variance (COV) distributions of surface curvature for input data (surface images). c, Surface curvature distributions for patient body structures sampled from each clustered group.
	\end{figure}

	DL applications in medical imaging are a promising research direction and can potentially lead to a paradigm shift in cancer diagnosis, prognosis, and radiotherapy. The proposed surface-to-volume framework offers a data integration solution to achieve image synthesizing between two image modalities without patients’ anatomical prior. The method can be extended to other volumetric image generation, such as magnetic resonance imaging and positron emission tomography. The proposed framework not only has the potential to enable real-time image-guided radiation therapy but also provide zero-dose synthetic CT images with anatomical details for lesion localization. However, cultivating trust in DL model safety and promoting innovative DL methods to manage risk\cite{RN720} are crucial for future clinical implementation, such as organ motion management, irregular breathing pattern recognition, and image-guided radiation therapy. Based on current literature\cite{RN727, RN728, RN729, RN730, RN719, RN718, RN725}, the robustness of DL models has become an area of active research. We classify the challenges and associated potential solutions in three categories: 1) data quality issues (bias or datasets limitation); 2) model form uncertainty (varying network architectures or different training strategies); 3) misleading evaluation (inadequate quantitative metrics). For 1) all the available and adequately evaluated data (for instance, relevant disease sites) should be ideally explored to serve as external validity\cite{RN718}. For 2) literature\cite{RN721, RN722} indicates that DL models with extensive layers can still be helpful for computer-aided detection problems even with limited training datasets. Meanwhile, unnecessarily complex models may be introduced because of the support to publication novelty\cite{RN718}. This complexity increases the difficulty of deploying or maintaining models in the clinic\cite{RN723}. In this work, we focus on patient-specific applications to ensure the relevancy of training data. The network performance is evaluated by using the relevant metrics to examine the absolute quality and structural similarity of the reconstructed volumetric images to the ground truth. If irregular patterns are detected in a specific patient, additional pre-operational quality assurance images can be used for training to extend the model’s applicability. We also keep the simplicity of the model architecture for interpretability\cite{RN367, RN280, RN724}.

	\bigbreak
	
	\noindent 
	\section{Conclusion}
	
	We have demonstrated that data-driven modeling can enable volumetric image generation using surface information with patient-specific priors (4DCT). The proposed surface-to-volume network can effectively correlate the hidden surface features to 3D patient anatomy and synthesize patient-specific CT images. The approach provides a potential integral solution via data integration for motion management in radiotherapy since the model can directly assimilate data without hardwired first-principal modeling of organ motion. In principle, the surface-to-volume image generation method can be extended to real-time imaging for radiosurgery, interventional procedures, or ultra-high dose rate FLASH radiotherapy. Such an imaging method does not make patients receive concomitant radiation doses due to imaging and can verify the treatment delivery.

	\noindent 
	\bigbreak
	{\bf ACKNOWLEDGEMENT}
	
	This research is supported in part by the National Institutes of Health under Award Number R01CA215718, R56EB033332, R01EB032680, and P30CA008748.

	\noindent 
	\bigbreak
	{\bf Disclosures}
	
	The authors declare no conflicts of interest.

	\noindent 
	
	\bibliographystyle{plainnat}  
	\bibliography{arxiv}      
	
\end{document}